\documentclass[twocolumn]{article}
\pdfoutput=1

\usepackage{arxiv}

\usepackage[utf8]{inputenc} 
\usepackage[T1]{fontenc}    
\usepackage{hyperref}       
\usepackage{url}            
\usepackage{booktabs}       
\usepackage{amsmath}
\usepackage{amsthm}
\usepackage{enumitem}
\setlist[itemize]{align=parleft,left=0pt..1em}
\setlist[enumerate]{align=parleft,left=0pt..1.5em,label=\large\protect\textcircled{\small\arabic*}}
\usepackage[font=scriptsize]{caption}
\usepackage{amsfonts}       
\usepackage{nicefrac}       
\usepackage{microtype}      
\usepackage{lipsum}		
\usepackage{graphicx,epstopdf}
\usepackage{natbib}
\usepackage{doi}
\usepackage{inconsolata}
\usepackage{xcolor}
\usepackage{listings}
\usepackage{booktabs} 

\usepackage{hyperref} 

\hypersetup{
    colorlinks=true,
    linkcolor=blue,
    filecolor=magenta,      
    urlcolor=cyan,
}

\lstdefinestyle{lua}{
  language=[5.3]Lua,
  basicstyle=\ttfamily,
  keywordstyle=\color{magenta},
  stringstyle=\color{blue},
  commentstyle=\color{black!50}
}
\lstdefinelanguage{Zencode}{
	morekeywords = {
		Given,
		When,
		Then,
		And,
		Scenario,
		Rule
	},
	sensitive=true,
	morecomment=[l]{\#},
	morestring=[b]'
}
\lstdefinestyle{zencode}{
  language=Zencode,
  backgroundcolor=\color{white},
  basicstyle=\ttfamily,
  keywordstyle=\color{blue},
  stringstyle=\color{magenta},
  commentstyle=\color{black!50}
}

\title{Reflow}

\author{
  Denis Roio \\
	Dyne.org \\
    \And
	Alberto Ibrisevic \\
  Dyne.org \\
    \And
  Andrea D'Intino \\
  Dyne.org
}

\renewcommand{\headeright}{Dyne.org}
\renewcommand{\undertitle}{Zero Knowledge Multi Party Signatures with Application to Distributed Authentication}
\renewcommand{\shorttitle}{\textit{Reflow}}

\hypersetup{
  pdftitle={Reflow multi-party signatures - Secure BLS multi-signatures Zero-knowledge proof },
  pdfsubject={cs.CR},
  pdfauthor={Denis Roio, Alberto Ibrisevic, Andrea D'Intino},
  pdfkeywords={crypto, sign, zero-knowledge, multi-party,
    material passport, authentication, valueflows, resource event agent},
}

\begin{document}

\twocolumn[
  \begin{@twocolumnfalse}

\maketitle

\begin{abstract}
Reflow is a novel signature scheme supporting unlinkable signatures by multiple parties authenticated by means of zero-knowledge credentials. Reflow integrates with blockchains and graph databases to ensure confidentiality and authenticity of signatures made by disposable identities that can be verified even when credential issuing authorities are offline.  We implement and evaluate Reflow smart contracts for Zenroom and present an application to produce authenticated material passports for resource-event-agent accounting systems based on graph data structures. Reflow uses short and computationally efficient authentication credentials and can easily scale signatures to include thousands of participants.
\end{abstract}

\vspace{2cm}

 \end{@twocolumnfalse}
]

\section{Introduction}

Multi-party computation applied to the signing process allows the
issuance of signatures without requiring any of the participating
parties to disclose secret signing keys to each other, nor requires the
presence of a trusted third-party to receive them and compose the
signatures. However, established schemes have shortcomings. Existing
protocols do not provide the necessary efficiency, re-randomization or
blind issuance properties necessary for the application to trust-less
distributed systems. Those managing to implement such privacy preserving
features are prone to rogue-key attacks \citep{ietf-bls} since they
cannot grant that signatures are produced by legitimate key holders.

The lack of efficient, scalable and privacy-preserving signature schemes
impacts distributed ledger technologies that support 'smart contracts'
as decentralized or federated architectures where trust is not shared
among all participants, but granted by one or more authorities through
credential issuance for the generation of non-interactive and unlinkable
proofs.

Reflow applies to the signature process a mechanism of credential
issuance by one or more authorities for the generation of
non-interactive and unlinkable proofs, resulting in short and
computationally efficient signatures composed of exactly two group
elements that are linked to each other. The size of the signature
remains constant regardless of the number of parties that are signing,
while the credential is verified and discarded after signature
aggregation. While being signed, duplicates may be avoided by collecting
unlinkable fingerprints of signing parties, as they would invalidate the
final result. Before being able to sign, a one-time setup phase is
required where the signing party collects and aggregates a signed
credential from one or more authorities.

Our evaluation of the Reflow functions shows very promising
results: basic session creation takes about $20ms$, while signing $73ms$ and
verification $40ms$ on average consumer hardware.

\subsection*{Overview}

Reflow provides a production-ready implementation that is easy to embed in end-to-end encryption applications. By making it possible for multiple parties to anonymously authenticate and produce untraceable signatures, its goal is to leverage privacy-by-design scenarios that minimize the information exchange needed for document authentication.

\begin{figure}
  \caption{Basic credential authentication }
  \label{fig:credential_diagram}
  \centering
  \includegraphics[width=0.5\textwidth]{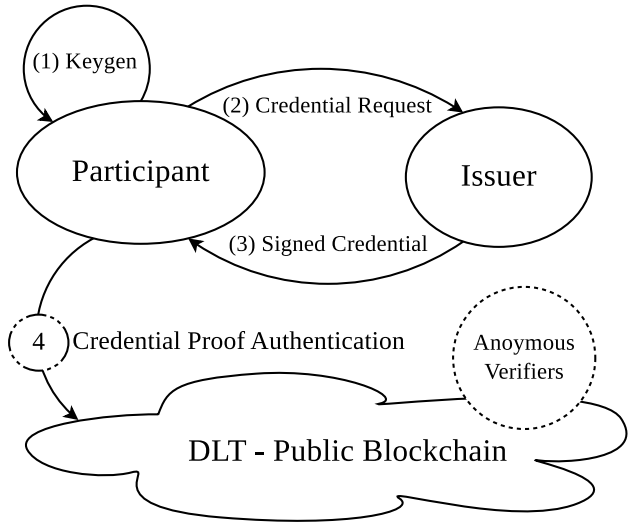}
\end{figure}

The participation to a signature will be governed by one or more issuers holding keys for the one-time setup of signature credentials. The steps outlined below are represented in figure \ref{fig:credential_diagram}:

\begin{enumerate} 
  \item Participant generates keys
  \item Participant sends a credential request to Issuer
  \item Issuer signs the credential request and sends it back
  \item Participant can create anonymous credential proofs
\end{enumerate}


Following this setup any participant will be able to produce a
zero-knowledge proof of possession of the credential, which can be
verified by anyone anonymously on the blockchain.

The base application of Reflow is the privacy preserving collective
signature of digital documents for which the signed credential proof is a
requirement to participate to any signature process.

A Reflow signature process is best described in 3 main steps:
session creation, signature and verification.

\paragraph*{1. Anyone creates a session}

\begin{figure}
  \caption{Session Creation}
  \label{fig:create_session}
  \centering
  \includegraphics[width=0.5\textwidth]{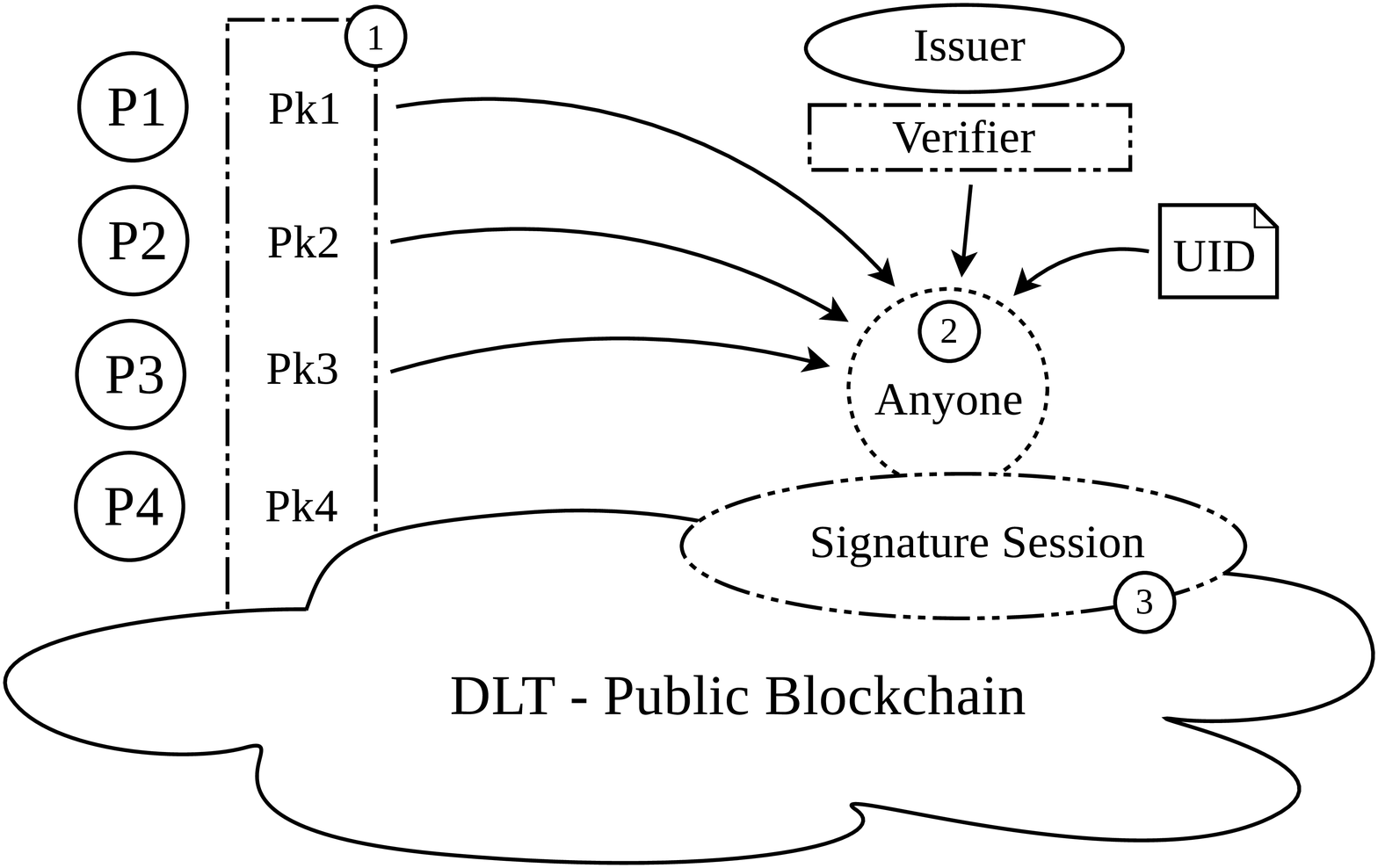}
\end{figure}

A session may be created by anyone, no credentials are required, but only information that should be public: the public keys of participants holding a credential to sign, the public verifier (public signature key) of the issuer who has signed the credentials and at last a document to be signed. The steps below are represented in figure \ref{fig:create_session} and illustrate how a signature session is created: 

\begin{enumerate}
  \item Participants will publish their public keys, available to anyone 
  \item Anyone may indict a signature session by selecting a document, the public keys of signing participants and the issuer
  \item The signature session is then published without disclosing the identity of any participant
\end{enumerate}

A Reflow signature session can then be published for verification and its existance may be confidentially communicated to the participants elected to sign it. The possession of the session will allow to disclose the identity of the participants, but only that of the Issuer and the document (or the hash of it) signed.

\paragraph*{2. Participants sign the session}

\begin{figure}
  \caption{Sign and Verify}
  \label{fig:verify_sign}
  \centering
  \includegraphics[width=0.5\textwidth]{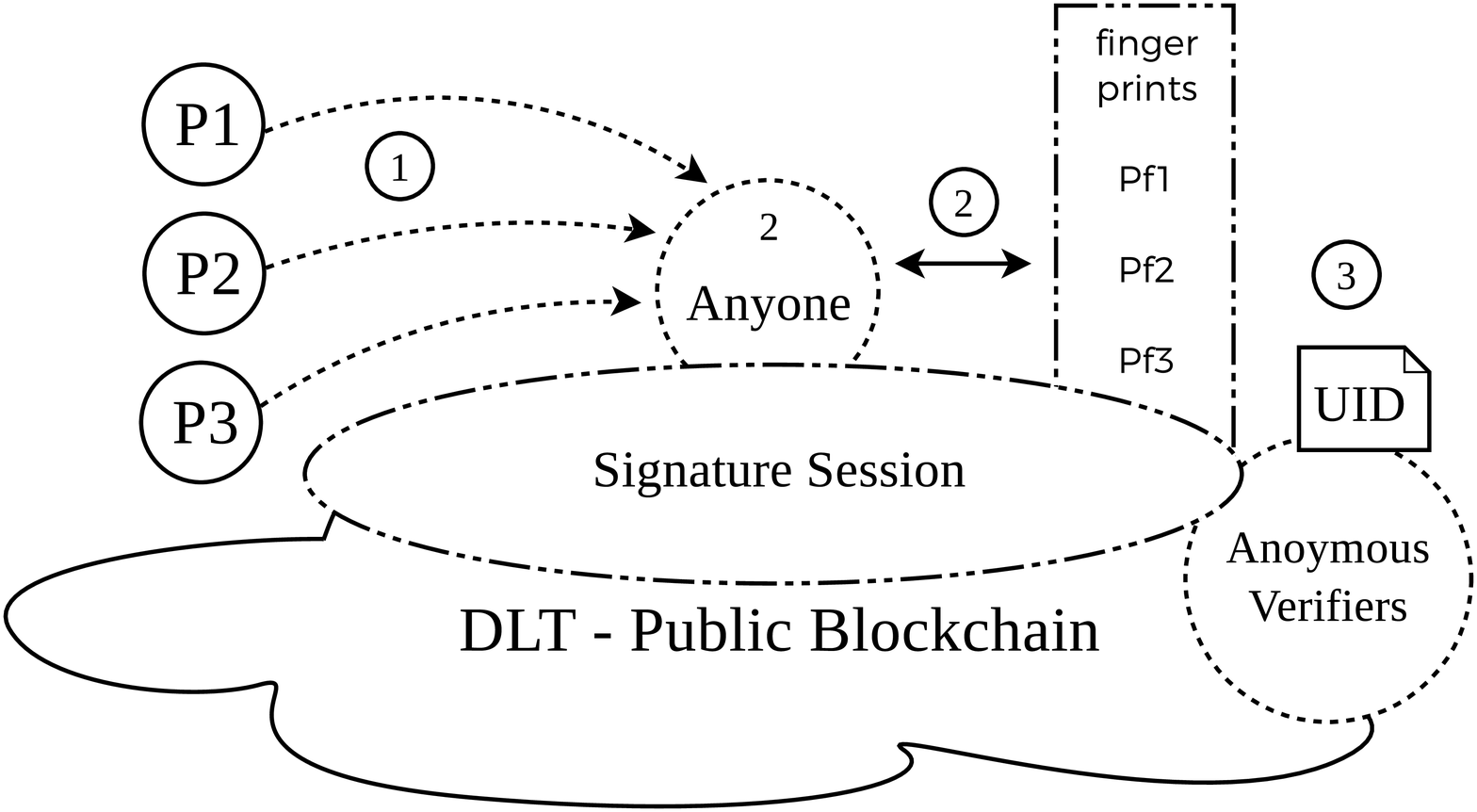}
\end{figure}

Only elected participants that were initially chosen to sign the
session may sign it, this is forced through credential
authentication. Whenever they know about the existance of a session
requesting their signature, they may chose to sign it. The steps below
are illustrated in figure \ref{fig:verify_sign}.

\begin{enumerate}
  \item Participants may be informed about the signature session and
    may create an anonymous signature to be added to the session 
  \item Anyone may check that the anonymous signature is authentic
    and not a duplicate, then adds it to the session
  \item Anyone may be informed about the signature session and be able to verify if the document is signed by all and only all participants
\end{enumerate}

A delicate aspect of BLS signatures is avoiding double-signing: if a
participant signs twice then the whole signature will never be valid.
Relying on a stateless credential authentication alone does not avoid
this case in Reflow, therefore we use a list of anonymous
"fingerprints" of the signature related to the document being signed.
Each signature will produce a participant fingerprint saved in a list
that is checked against duplication before adding a new
signature. This procedure adds significant computational overhead for
sessions with a large number of participants, but it can be switched
off in a system that avoids double-signing in its own architecture.

\paragraph*{Anyone verifies signatures}

Until the session will have collected all the signatures of
participants, its verification will not be valid. It is also
impossible to know if all participants have signed the session or how
many are missing. Anyone can verify the state of the signature session
in any moment just by having the document and the session, as
illustrated in figure \ref{fig:verify_sign} along with the signature
process.

Configurable features may be introduced in the Reflow signature
flow that may or may not disclose more information, for instance who
has indicted the signature session, what documents are linked to
signatures and how many participants were called to sign: this depends
from the implementation and the metadata it may add to signature
sessions or the communication protocols adopted.  In any case the
basic signature and verification flow of Reflow requires that
only one identity is really made public and is that of the issuer.

\subsection*{Applications}

Moving further in envisioning the possibilities opened by Reflow
is important to state the possibility to aggregate (sum) signatures into
compact multi-signatures, a core feature of our BLS based signature
scheme \citep{compact-multisig}.

\paragraph*{Need to Know.}

The base implementation exploiting this feature is that of a signature
scheme for a single document split in separate sections to be signed by
different participants: all the signatures can be later aggregated in a
single one proving the whole document has been signed by all
participants, without being disclosed to all of them in its entirety.
This application helps to enforce the principles of need-to-know and
least privilege to the access of information \citep{info-protection} and
is useful for the realization of privacy-aware applications in various
sectors, for instance for medical and risk mitigation analysis.

\paragraph*{Disposable Identities}

A Disposable Identity is based on four properties common to other
authentication and identification systems: verifiability, privacy,
transparency, trustworthiness; then in addition introduces a fifth
property: disposability. Disposability permits purpose-specific and
context-driven authentication, to avoid linking the same identity across
different authentication contexts \citep{dispid}.

Reflow can be adopted by such an application to remove the need of
context-free identifiers and implement authentication functions through
a disposable identity whose traceability is bound to a context UID.
Furthermore, being a signature scheme, Reflow can add the feature of
context-free verifiable signatures (and multi-signatures) that are
untraceable in public, but can be traced and even revoked in a known
context by the means of context-specific signature fingerprints.

This scenario is relevant for the implementation of privacy-preserving
public sector applications that allow authentication and signatures
through disposable identity systems \citep{dispid-kranenbu}.

\paragraph*{Material Passport.}
Drawing on feature of multiple signature aggregation, Reflow can
be used to implement a {material passport} for circular economy
applications \citep{material-passport} to maintain the genealogy of a
specific product, providing authenticated information about  the whole
set of actors, tools, collaborations, agreements, efforts and energy
involved in its production, transportation and disposal
\citep{Reflow-os}.

The provision of the information that forms the content of the material
passport should be done by every actor in the supply chain and among the
most important technical necessities for such an application are the
confidentiality issues regarding access to information and the
guarantees of the quality of information \citep{resources-passport}.

As an ideally simple and effective ontology we adopt a
Resource-Event-Agent model \citep{REA} and the ValueFlows vocabulary
\citep{valueflows} to organize knowledge in a graph made of 3 main
type of nodes:

\begin{enumerate}
  \item Events that are the means of creation and transformation of Resources
  \item Agents capable of creating Events and Processes
  \item Processes where more than one Agent can consume and create Events
\end{enumerate}

\begin{figure}
  \caption{Valueflows}
  \label{fig:valueflows}
  \centering
  \includegraphics[width=0.5\textwidth]{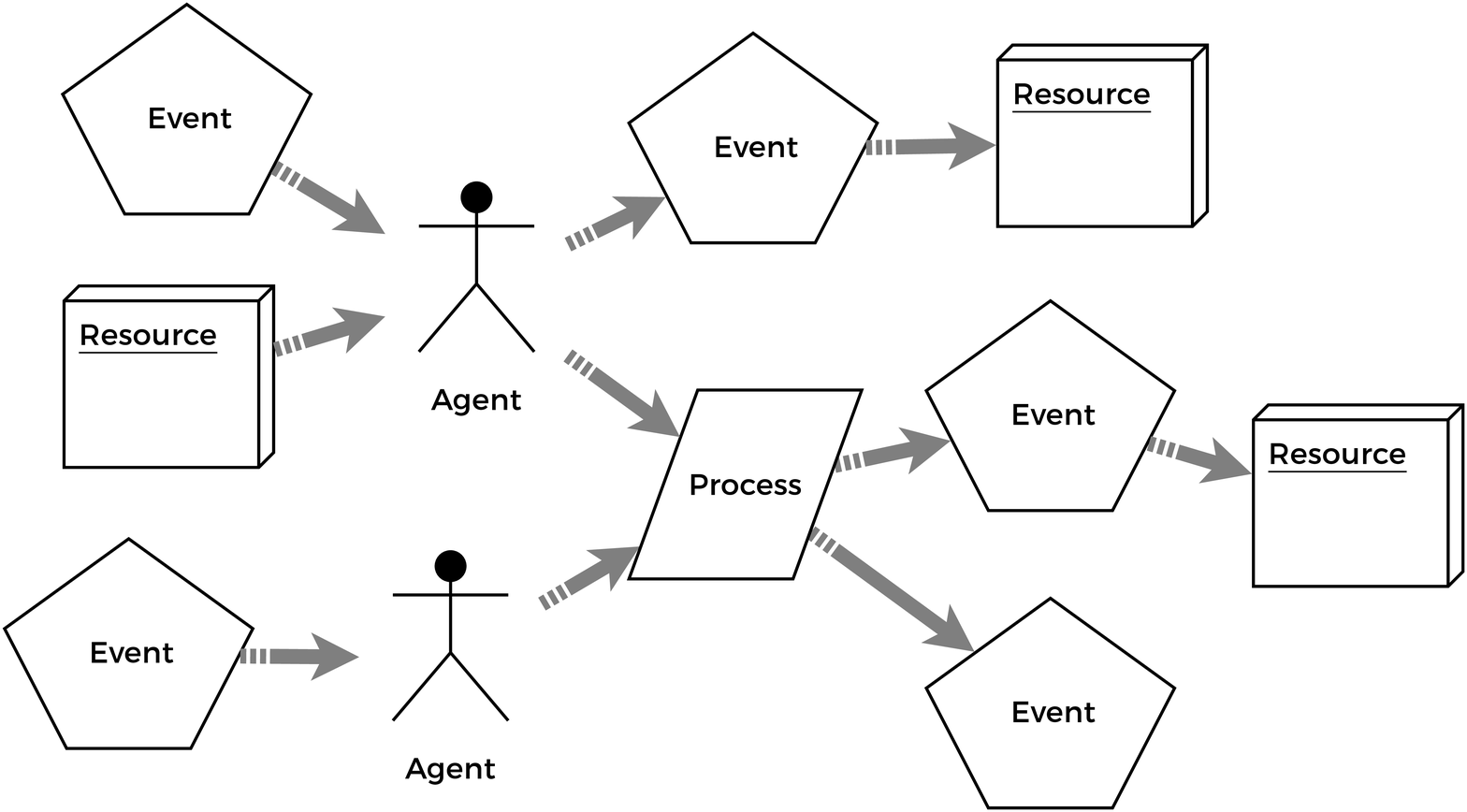}
\end{figure}

We then consider Resources as material passports made of the track and
trace of all nodes - Events, Agents and Processes - they descend from.
The  material passport is an authenticated graph structure: in figure
\ref{fig:valueflows} the Resources on the right side have an UID which
is the aggregation of all UIDs of elements leading to their existance.

The integrity of the material passport can be verified by recalculating
the UID aggregation and see it matches the signed one attached to the
Resource. In case one or more UIDs are wrong or missing, the Resource
will not verify as valid. In brief is possible that:

\begin{enumerate}
  \item One or more Agents may interact to create material passports
  \item A material passport is the aggregation of all parent nodes
  \item Anyone may verify the integrity and validity of any material passport
  \item The verification of a material passport does not reveal the identity of Agents contained
  \item One may export and import a material passport as a graph query
\end{enumerate}

Reflow's unlinkability of credentials and signatures satisfy the privacy
requirement for the material passport, while the possibility to
aggregate and link all the elements of its graph allow to group multiple
signatures into a single compact one, without requiring any interaction
with the previous signers. The material passport signature will then be
the sum of all Agents, Processes and Events involved, created or
consumed for it. The Reflow material passport is the authenticated,
immutable and portable track record of all nodes connected in its graph:
material passports can be signed on export and verified on import, which
makes them reliably portable from one graph to another in a federated
environment.

\subsection*{This paper makes four key contributions}

\begin{itemize}
\item We describe the signature scheme underlying Reflow, including how
  key generation, signing and verification operate (Section
  \ref{sec:signature}). The scheme is an application of the BLS signature scheme
  \citep{asiacrypt-bls} fitted with features to grant the unlinkability of
  signatures and to secure it against rogue-key attacks.
\item We describe the credential scheme underlying Reflow, including how
  key generation, issuance, aggregation and verification of credentials operate
  (Section \ref{sec:credential}). The scheme is an application of the Coconut
  credential scheme \citep{coconut-2018} that is general purpose and can be
  scaled to a fully distributed issuance that is re-randomizable.
\item We implement a Zencode scenario of Reflow to be executed on and
  off-chain by the Zenroom VM, complete with functions for public credential
  issuance, signature session creation and multi-party non-interactive signing
  (Section \ref{sec:implementation}). We evaluate the performance and cost of
  this implementation on on-site and on-line platforms leveraging end-to-end
  encryption (Section \ref{sec:evaluation}).
\item We implement an efficient end-to-end encryption scheme for the
authentication of graph data structures as observed in the material
passport use-case. The scheme allows to simplify the complexity of track-and-trace implementations by making it sufficient to proceed by one level of depth and verify the integrity of aggregated signatures at each step.
\end{itemize}

\subsection*{Notations and assumptions}

We will adopt the following notations:
\begin{itemize}
    \item $\mathbb{F}_p$ is the prime finite field with $p$ elements (i.e. of prime order $p$); 
    \item $E$ denotes the (additive) group of points of the curve BLS-383 \citep{bls383} which can be described with the Weierstrass form  $y^2=x^3 + 16$; 
    \item $E_T$ represents instead the group of points of the twisted curve of BLS-383, with embedding degree $k=12$. The order of this group is the same of that of $E$;
\end{itemize}
We also require defining the notion of a cryptographic pairing. Basically it is a function $e: \mathbb{G}_1\times\mathbb{G}_2\to \mathbb{G}_T$, where $\mathbb{G}_1,\mathbb{G}_2$ and $\mathbb{G}_T$ are all groups of same order $n$, such that satisfies the following properties:
\begin{itemize}
    \item [i.] \emph{Bilinearity}, i.e. given $P_1,Q_1\in\mathbb{G}_1$ and $P_2,Q_2\in\mathbb{G}_2$, we have 
    \begin{align*}
        e(P_1+Q_1,P_2) = e(P_1,P_2)\cdot e(Q_1,P_2) \\
        e(P_1,P_2+Q_2) = e(P_1,P_2)\cdot e(P_1,Q_2)
    \end{align*}
    \item[ii.] \emph{Non-degeneracy}, meaning that for all $g_1\in\mathbb{G}_1, g_2\in\mathbb{G}_2$, $e(g_1,g_2)\ne 1_{\mathbb{G}_T}$, the identity element of the group $\mathbb{G}_T$;
    \item[iii.] \emph{ Efficiency}, so that the map $e$ is easy to compute;
    \item[iv. ] $\mathbb{G}_1\ne \mathbb{G}_2$, and moreover, that there exist no efficient homomorphism between $\mathbb{G}_1$ and $\mathbb{G}_2$.
\end{itemize}
For the purpose of our protocol we will have $\mathbb{G}_1 = E_T$ and $\mathbb{G}_2 = E$, and $\mathbb{G}_T\subset \mathbb{F}_{p^{12}}$ is the subgroup containing the $n$-th roots of unity, where $n$ is the order of the groups $E$ and $E_T$. Instead $e: E_T  \times E\to \mathbb{G}_T$ is the \emph{Miller pairing}, which in our work is encoded as the method \verb!miller(ECP2 P, ECP Q)!. \\
To conclude, the credential scheme in section \ref{sec:credential} uses non-interactive zero-knowledge proofs (for short NIZK proofs) to assert knowledge and relations over discrete logarithm values. They will be represented using the notation introduced by \cite{camenisch} as 
\[
\text{NIZK}\{(x,y,\dots): text{statements about x, y,}\dots \}
\]

\section{Signature}
\label{sec:signature}

A \emph{BLS signature} is a signature scheme whose design exploits a cryptographic pairing. As for other well known algorithm such as ECDSA, it will work following these three main steps:
\begin{itemize}
    \item  \textbf{Key Generation phase.} For a user who wants to sign a message $m$, a secret key $sk$ is randomically chosen uniformly in $\mathbb{F}_n$, where $n$ is the order of the groups $\mathbb{G}_1, \mathbb{G}_2, \mathbb{G}_T$. The corresponding public key $pk$ is the element $sk\cdot G_2\in E_T$;
    \item   \textbf{Signing phase.} The message $m$ is first hashed
      into the point $U\in E$, which in our scheme is done by the
      method \verb!hashtopoint!; the related signature is then given
      by $\sigma = sk\cdot U$;
    \item   \textbf{Verification phase.} For an other user that wants to verify the authenticity and the integrity of the message $m$, it needs to
    \begin{itemize}
        \item [1.] parse $m, pk$ and $\sigma$
        \item [2.] hash the message $m$ into the point $U$ and then check if the following identity holds,
        \[
        e(pk,U) = e(G_2,\sigma)
        \]
    \end{itemize}
\end{itemize}
If verification passes it means that $\sigma$ is a valid signature for $m$ and the protocol ends without errors.
\begin{proof}
 [Proof of the verification algorithm:] By using the definitions of the elements involved and exploiting the property of the pairing $e$ we have
\[
\begin{split}
    e(pk,U) &= e(sk\cdot G_2, U) \\
            &= e(G_2,U)^{sk}\\
            &= e(G_2,sk\cdot U)\\
            &= e(G_2,\sigma)
\end{split}
\]
\end{proof}

BLS signatures present some interesting features. For instance, the length of the output $\sigma$ competes to those obtained by ECDSA and similar algorithms; in our specific case, by using BLS-383 \citep{bls383}, it will be 32 Bytes long, which is typically a standard nowadays. Then, since this curve is also pairing-friendly, meaning that (with the assumption made on $e$) signature and verification are obtained in very short time. Moreover, BLS supports also aggregation, that is the ability to aggregate a collection of multiple signatures $\sigma_i$ (each one related to a different message $m_i$) into a singular new object $\sigma$, that can be validated using the respective public keys $pk_i$ in a suitable way. This is possible thanks to the fact that $\sigma_i\in G_1 \forall i$, giving to the algorithm an homomorphic property. We will show now how this last feature can be attained in the context of a multi-party computation using the same message $m$ but different participants.

\subsection*{Session Generation}

After the key generation step we introduce a new phase called \textbf{session generation}, where the signature is initialized; anyone willing to start a signing session on a message $m$ will create:
\begin{itemize}
    \item[1.] a random $r$ and its corresponding point $R = r\cdot G_2$
    \item[2.] the sum of $R$ and all $pk$ supposed to participate to the signature such as $P = R + \sum_i pk_i$
    \item[3.] the unique identifier \verb!UID! of the session calculated as hash to point of the message $m$, such as $U = H(m)\in E$, where $H$ is a combination of a cryptographic hash function (treated as a random oracle) together with an encoding into elliptic curve points procedure
    \item[4.] the first layer of the signature $\sigma \leftarrow r\cdot U$, later to be summed with all other signatures in a multi-party computation setup resulting in the final signature as $\sigma \leftarrow r\cdot U + \sum_i sk_i\cdot U$
    \item[5.] the array of unique fingerprints $\zeta_i$ of each signature resulting from the credential authentication (see section \ref{sec:credential}) 
\end{itemize}
After this phase is terminated, every participants involved in the session start their own signing phase during the session, producing (from the same message $m$) their respective $\sigma_i$'s. The final signature $\sigma$ is then computed in this way: first of all let us call $\sigma_0 = r\cdot U$, then supposing $k$ participants have already aggrgated their $\sigma_i$, obtaining a partial signature $S_k$, the $(k+1)$-th one will compute
\[
  S_{k+1}=S_k+\sigma_k = \sum_{i=0}^{k+1} \sigma_i
\]
Finally, the resulting output will be $\sigma = S_N$, where $N$ is the total number of the signers of the session. In order to verify that $\sigma$ is valid we compute $P = R + \sum_{i=0} ^N pk_i$, where $R = r\cdot G_2$, working as a public key with respect to the nonce $r$, which instead is kept secret. Verification is then performed by checking if the following identity holds
    \[ 
    e(P,U) = e(G_2,\sigma)
    \]
If verification passes without errors it means that $\sigma$ is a valid aggregated signature of $m$.
\begin{proof}
By recalling that $\sigma = r\cdot U + \sum_{i=1}^N \sigma_i$, $P = R + \sum_{i=1}^N pk_i$, by using the property of the pairing $e$ we have
\[
\begin{split}
    e(P,U)  &= e(R + \sum_{i=1}^N pk_i, U) \\
            &= e((r + \sum_{i=1}^N sk_i)\cdot G_2,U)\\
            &= e(G_2,U)^{r + \sum_{i=1}^N sk_i}\\
            &= e(G_2,(r + \sum_{i=1}^N sk_i)\cdot U) \\
            &= e(G_2,r\cdot U + \sum_{i=1}^N \sigma_i) \\
            &= e(G_2,\sigma)
\end{split}
\]
\end{proof} 
We conclude this section with a final consideration on this feature. We recall that in the generation of the aggregated signature $\sigma$ we used as a starting point the variable $\sigma_0 = r\cdot U$, but in literature it is also common to find instead simply the base point $G_2$. The choice of randomizing it (providing that the random number generator acts as an oracle) helps in preventing replay attacks, since the signature generated by the process is linked to the session in which is produced, for if an attacker managed to get some information from $\sigma$, it would be difficult to use it in order to forge new signatures.

\section{Credential}
\label{sec:credential}

Following the guidelines of Coconut, the credentials issuing scheme works as follows:
\begin{itemize}
\item [1.] the issuer generates its own keypair $(s_k,v_k)$, where $s_k=(x,y)\in\mathbb{Z}^2$ is the pair of secret scalars (the signing key) and $v_k=(\alpha, \beta)=(x\cdot G_2,y\cdot G_2)$ is the verifying key, made by the related pair of public points over $E_T$; 
\item [2.] the user $i$, with its respective keys $(sk_i, PK_i)$ make a credential request on its secret attribute $ck_i\in\mathbb{Z}$ to the issuer, represented by $\lambda$ which contains a zero-knowledge proof $\pi_s$ of the authenticity of user $i$; 
\item[3.] the issuer, after having received $\lambda$, verifies the proof $\pi_s$ at its inside, and if it passes, then releases to user $i$ a credential $\tilde{\sigma}$ signed used its own key $sk$.
\end{itemize}
Step 1. is self-explanatory. Steps 2. and 3. require a bit more effort, in fact in order to build a valid request $\lambda$, and so also a valid proof $\pi_s$, first of all the user must produce an hash digest for the attribute $ck_i$, that we call $h$, then computes two more variables $c$ and $s_h$ defined as
\begin{align*}
c &= r\cdot G_1 + h\cdot HS \\
s_h &= (a,b) = (k \cdot G_1, k\cdot \gamma + h\cdot c)
\end{align*}
where $r$ and $k$ are fresh randomly generated integers, $HS$ is an hard-encoded point on the curve $E$, and $\gamma = ck_i\cdot G_1$. These two variables are alleged in the credential request $\lambda$ produced in \verb!prepare_blind_sign! and are needed to the verifier to assure the authenticity of the user through the proof $\pi_s$, which requires as input $h, k, r, c$. The Non-Iteractive Zero Knowledge proof (for short NIZK proof) $\pi_s$ generated by the function \verb!blind_sign! is computed as follows:
\begin{itemize}
    \item \textbf{Randomization phase.} Three new nonces $w_h, w_k, w_r\in \mathbb{Z}$ are generated, each one related to the input variables $h, k, r$ respectively as we will show soon; 
    \item \textbf{Challenge phase.} The protocol creates three commitment values, namely $A_w, B_w, C_w$ defined as follows
    \begin{align*}
        A_w &= w_k\cdot G_1 \\ 
        B_w &= w_k\cdot\gamma + w_h\cdot c \\
        C_w &= w_r\cdot G_1 + w_h \cdot HS
    \end{align*}
    Then these variables are used as input of a function $\varphi$ producing an integer $c_h=\varphi(\{c,A_w,B_w,C_w\})$;
    \item \textbf{Response phase.} In order that the proof can be verified the protocol generates three more variables which are alleged inside the proof itself and link the nonces $w_h, w_k, w_r$ with $h, k, r$, i.e:
    \begin{align*}
        r_h &= w_h - c_h h \\
        r_k &= w_k - c_h k \\
        r_r &= w_r - c_h r
    \end{align*}
\end{itemize}
So basically the proof $\pi_s$ contains the three response variables $r_h, r_k, r_r$ and also the commitment value $c_h$, that can be used for a predicate $\phi$ which is true when computed on $h$. Once the verifier receives the request $\lambda$, in order to check if the proof is valid it should be able to reconstruct  $A_w, B_w, C_w$ by doing these computations,
\begin{align*}
\widehat{A}_w &= c_h\cdot a + r_k\cdot G_1 \\
\widehat{B}_w &= c_h\cdot b + r_k\cdot \gamma + r_h\cdot c \\
\widehat{C}_w &= c_h\cdot c + r_r\cdot G_1 + r_h\cdot HS
\end{align*}
If the request is correct, then we will have that 
\begin{equation}\label{challenge pi_s}
\varphi(\{c,\widehat{A}_w,\widehat{B}_w,\widehat{C}_w\}) = \varphi(\{c,A_w,B_w,C_w\}) = c_h
\end{equation}
and verification is thus complete, meaning that the verifier has right to believe that the prover actually owns the secret attribute $ck_i$ associated to the public variable $\gamma$ and that consequently has produced a valid commitment $c$ and (El-Gamal) encryption $s$; in other words,
\begin{align*}
\pi_s = \text{ NIZK}\{&(ck_i, h, r, k): \\
&\gamma = ck_i\cdot G_1, \\
&c = r\cdot G_1 + h\cdot HS, \\
&s_h = (k \cdot G_1, k\cdot \gamma + h\cdot c), \\
&\phi(h)=1\}
\end{align*}
At this point the user will now have a blind credential $\tilde{\sigma} = (c, \tilde{a}, \tilde{b})$ issued by the authority, where
\begin{align*} \\
\tilde{a} &= y \cdot a \\
\tilde{b} &= x\cdot c + y\cdot b
\end{align*}
The user then will have to un-blind it using its secret credential key, obtaining $\sigma_{ck} = (c, s) = (c, \tilde{b} - ck_i (\tilde{a}))$, which will use to prove its identity when signing a message. The procedure is similar to the one seen before with some extra details:
\begin{itemize}
    \item \textbf{Setup.} As for the the BLS signature, an elliptic point $U$, associated to the hash of the message to sign, is required as an Unique Identifier (\verb!UID!) for the signing session;
    \item \textbf{Credential proving.} The user produces two cryptographical objects $\theta$ (containing a new proof $\pi_v$) and $\zeta$ (which is unequivocally associated to $U$) through \verb!prove_cred_uid!, taking as input its own credential $\sigma$, the related secret attribute $c_k$, the authority public key $v_k = (\alpha, \beta)$ and the session point $U$.
\end{itemize}
The new objects $\theta$ and $\zeta$ are derived as follows:
\begin{itemize}
    \item as before the user hashes $ck$ into $h$, and this time generates two random values $r$ and $r'$;\\
    \item next, it randomizes its credential $\sigma_{ck}$ into $\sigma_{ck}' = (c', s') = (r'\cdot c, r'\cdot s)$ and then computes two elliptic curve points $\kappa$ and $\nu$ as
    \begin{align*}
        \kappa &= \alpha + h\cdot\beta + r\cdot G_2\\
        \nu &= r \cdot c'
    \end{align*}\\
    \item finally, $\theta$ will be the t-uple $(\kappa, \nu, \pi_v, \phi')$, where $\pi_v$ is a valid zero knowledge proof of the following form
    \begin{align*}
        \pi_v = \text{ NIZK}\{&(h, r): \\
        &\kappa = \alpha + h\cdot\beta + r\cdot G_2\\
        &\nu = r \cdot c' \\
        &\phi'(h)=1\}
    \end{align*}
    with $\phi'$ being a predicate which is true on $h$; 
    \item $\zeta$ will be instead the elliptic curve point obtained as $h\cdot U \in E$.
\end{itemize}
Building the proof $\pi_v$ requires similar steps as seen for $\pi_s$, in fact we create three commitment values $A_w, B_w, C_w$ defined as
\begin{align*}
    A_w &= \alpha + w_h\cdot \beta + w_r \cdot G_2 \\
    B_w &= w_r\cdot c' \\
    C_w &= w_h\cdot U
\end{align*}
where $w_h, w_r$ are fresh generated nonces; then we set the challenge as the computation of $c_h=\varphi(\{\alpha,\beta,A_w,B_w,C_w\})$ with the related responses 
\begin{align*}
    r_h &= w_h - hc \\
    r_r &= w_r - rc
\end{align*}
The values of $r_h$ and $r_r$ are stored inside $\pi_v$ which will be then sent through $\theta$ (together with $\zeta$) from the prover to the verifier. In order to check that the user has legitimately generated the proof and at the same time is the owner of the credential the following steps must be made:
\begin{itemize}
    \item[1.] extracting $\kappa, \nu, \pi_v$ (which is $(r_h,r_r, c_h))$ from $\theta$;
    \item[2.] reconstructing the commitments $A_w, B_w, C_w$ as
    \begin{align*}
        \widehat{A}_w &= c_h\cdot \kappa + r_r\cdot G_2 + (1 - c_h) \cdot\alpha + r_h \cdot \beta \\
        \widehat{B}_w &= r_r\cdot c' + c_h\cdot \nu \\
        \widehat{C}_w &= r_h\cdot U + c_h\cdot \zeta
    \end{align*}
    \item[3.] checking either that
    \begin{multline}\label{challenge pi_v}  \varphi(\{\alpha,\beta,\widehat{A}_w,\widehat{B}_w,\widehat{C}_w\}) = \\ \varphi(\{\alpha,\beta,A_w,B_w,C_w\}) = c_h,
    \end{multline}
    that $c' \ne O$, the point at infinity, and that
    \begin{equation}\label{miller}
    e(\kappa, c') = e(G_2, s' + \nu)
    \end{equation}
    \end{itemize}
Actually the predicate $\phi$ in the definition of $\pi_v$ can be thought as performing steps 2. and 3. and, if any of these fails the protocol will abort returning a failure, otherwise verification passes and the user can finally produce the signature.

\subsection*{Proof of the verification algorithms.}

We now show for the proof $\pi_s$ that actually by using the responses $r_h, r_k$ and $r_r$, together with $c_h$ and the other parameters inside $\lambda$, i.e. $s=(a,b), c$ and $\gamma$, using also the hard coded point $HS$, it is possible to reconstruct the commitments $A_w, B_w, C_w$:
\begin{align*}
    \widehat{A}_w   &= c_h\cdot a + r_k\cdot G_1 = c_h k \cdot G_1 + r_k \cdot G_1 \\
                    &= (c_h k + r_k)\cdot G_1 = (c_h k + w_k - c_h k)\cdot G_1 \\
                    &= w_k\cdot G_1 = A_w \\
    \widehat{B}_w   &= c_h\cdot b + r_k\cdot \gamma + r_h\cdot c \\
                    &= c_h\cdot (k\cdot \gamma + h\cdot c) + r_k\cdot\gamma + r_h\cdot c \\
                    &= (c_h k + r_k) \cdot\gamma + (c_h h + r_h\cdot c) \\
                    &= (c_h k + w_k - c_h k ) \cdot\gamma + (c_h h + w_h - c_h h)\cdot c \\
                    &= w_k\cdot\gamma + w_h\cdot c = B_w \\
    \widehat{C}_w   &= c_h\cdot c + r_r\cdot G_1 + r_h\cdot HS \\
                    &= c_h\cdot (r\cdot G_1 + h\cdot HS) + r_r\cdot G_1 + r_h\cdot HS \\
                    &= (c_h r + r_r)\cdot G_1 + (c_h h + r_h)\cdot HS \\
                    &= (c_h r + w_r - c_h r)\cdot G_1 + (c_h h + w_h - c_h h)\cdot HS \\
                    &= w_r \cdot G_1 + w_h\cdot HS = C_w
\end{align*} 
Regarding the second proof $\pi_v$, we have to prove both the identities (\ref{challenge pi_v}) and (\ref{miller}) hold. We will focus only on the latter since the former requires a similar approach on what we have have done for $\pi_s$, but with different parameters involved ($\kappa, \nu$, etc.). The left-hand side of the relation can be expressed as
\begin{align*}
    e(\kappa, c') &= e(\alpha + h\cdot\beta + r\cdot G_2, c') \\
    &= e(x\cdot G_2 + hy\cdot G_2  + r\cdot G_2,\tilde{r}\cdot G_1) \\
    &= e((x+hy+r)\cdot G_2, \tilde{r}\cdot G_1) \\
    &= e(G_2,G_1)^{(x+hy+r)\tilde{r}}
\end{align*}
using the substitution $c'=\tilde{r}\cdot G_1$, with $\tilde{r}\in\mathbb{F}_p$ since we know that $c'\in E$. For the right-hand side we have instead
\begin{align*}
    e(G_2, s' + \nu) &= e(G_2, r'\cdot s + r \cdot c') \\
    &= e(G_2, r'(\tilde{b} - ck_i y (\tilde{a})) + r \cdot c') \\
    &= e(G_2, r'(x\cdot c + y\cdot b - ck_i y\cdot a) + r \cdot c') \\
    &= e(G_2, r'(x\cdot c + y\cdot b - ck_i y\cdot a + r \cdot c)) 
\end{align*}
The second argument of the pairing can be rewritten as
\[
\begin{split}
    &r'(x\cdot c + y\cdot b - ck_i y\cdot a + r \cdot c) = \\
    &r'(x\cdot c + y(k\cdot \gamma + h\cdot c) - (ck_i) yk\cdot G_1 + r \cdot c) = \\
    &r'(x\cdot c + yh\cdot c + yk(ck_i)\cdot G_1  - yk(ck_i)\cdot G_1 + r \cdot c) = \\
    &r'(x + yh + r) \cdot c
\end{split}
\]
So, at the end
\[
\begin{split}
    e(G_2, s' + \nu) &= e(G_2, r'(x\cdot c + y\cdot b - dy\cdot a + r \cdot c)) \\
    &= e(G_2, r'(x + yh + r) \cdot c) \\
    &= e(G_2,(x+hy+r)\tilde{r}\cdot G_1) \\
    &= e(G_2,G_1)^{(x+hy+r)\tilde{r}}
\end{split}
\]
and (\ref{miller}) is finally proved. 
\qed

\subsection*{Security considerations.} 

As mentioned in Coconut \citep{coconut-2018}, BLS signatures and the
proof system obtained with credentials are considered secure by assuming
the existence of random oracles \citep{random-oracle}, together with the
decisional Diffie-Hellman Problem (DDH) \citep{DDH-problem}, the
external Diffie-Hellman Problem (XDH), and with the
Lysyanskaya-Rivest-Sahai-Wol Problem (LRSW) \citep{lrsw-assumption},
which are connected to the Discrete Logarithm. In fact, under these
assumptions, we have that our protocol satisfies unforgeability,
blindness, and unlinkability.

Reserves can be made about the maturity of pairing-based ellipcic
curve cryptography despite various efforts to measure its security and
design curve parameters that raise it, it is reasonable to consider
this as a pioneering field of cryptography in contrast to well tested
standards.

In addition to considerations on the maturity of EC, the future growth
of quantum-computing technologies may be able to overcome the Discrete
Logarithmic assumptions by qualitatively different computational
means. Reflow then may be vulnerable to quantum-computing
attacks, as well hard to patch, because the pairing-based design sits
at its core with the adoption of ATE / Miller loop pairing of curves
in twisted space, a practice that is not covered by research on
quantum-proof algorithms and will eventually need more time to be
addressed; however this is all speculative reasoning on what we can
expect from the future.

The Reflow implementation we are presenting in this paper and
that we have published as a Zenroom scenario ready to use is based on
the BLS383 curve \citep{bls383} that in the current implementation
provided by the AMCL library has shown to pass all lab-tests regarding
pairing properties, a positive result that is not shared with the
slightly different BLS381-12 curve adopted by ETH2.0. Debating the
choice of BLS381 is well beyond the scope of this paper, but is worth
mentioning that our lab-tests have proved also the BLS461 curve to work
in Reflow: it is based on a 461 bit prime and hence upgrades our
implementation to 128 bit security \citep{updating-key-pairings} against
attacks looking for discrete logs on elliptic curves
\citep{discrete-log-attack}.

At last the complexity and flexibility of Reflow in its
different applications, its optional use of fingerprint lists,
multiple UIDs saved from aggregation and other features covering the
different applications also represent a security risk in the technical
integration phase. We believe that the adoption of Zenroom and the
creation of a Zencode scenario addresses well this vulnerability by
providing an easy to use integrated development environment
(apiroom.net) and a test-bed for the design of different scenarios of
application for Reflow that can be deployed correctly granting
end-to-end encryption and data minimization according to privacy by
design guidelines \citep{privbydesign}.

\section{Implementation}
\label{sec:implementation}


In this section we illustrate our implementation of Reflow
keygen, sign and verify operations outlining for each:

\begin{itemize}
  \item the communication sequence diagram when necessary
  \item the Zencode statements to operate the transformation
\end{itemize}

Zencode is actual code executed inside the Zenroom VM, behind its
implementation the algorithms follow very closely the mathematical
formulation explained in this article and can be reviewed in the free
and open source code published at Zenroom.org. The Reflow Zencode
scenario implementation is contained inside the Zenroom source code.

To execute an example flow on one's own computer it is enough to visit
the website ApiRoom.net and select the Reflow scenario examples,
pressing play it will execute zencode and show input and outputs in
JSON format.

\subsection*{Credential Setup}

The Credential Setup sequence which has to be executed only once to
enable participants to produce signatures. This sequence is briefly
illustrated with a diagram and it consists in the creation of keypair
for both the Issuer, who will sign the participant credentials, and
the participant who will request an issuer credential.

\begin{lstlisting}[style=zencode,caption={Issuer Keygen}]
Given I am 'The Issuer'
When I create the issuer key
Then print my 'keys'  
\end{lstlisting}

Executed by the Zencode utterance:

\textbf{When I create the issuer key}

It will create a new \emph{issuer key} that can be used to sign
each new \emph{credential request}. Its public member \emph{.verify}
should be public and know to anyone willing to verify the credentials
of signers. It can be safely separated from the secret key and
exported with:

\begin{lstlisting}[style=zencode,caption={Issuer verifier}]
Scenario credential: publish verifier
Given that I am known as 'The Issuer'
and I have my 'keys'
When I create the issuer public key
Then print my 'issuer public key'
\end{lstlisting}

\textbf{Credential signature:} Generate a credential request and have
it signed by an Issuer, as well generate all the keys used to sign
documents. This procedure will generate private keys that should not
be communicated, as well BLS keys whose public section should be
communicated and can be later aggregated to create a seal (signature
session) and verify one.

\begin{figure}
  \caption{Keygen process sequence diagram}
  \label{fig:keygen}
  \centering
  \includegraphics[width=0.5\textwidth]{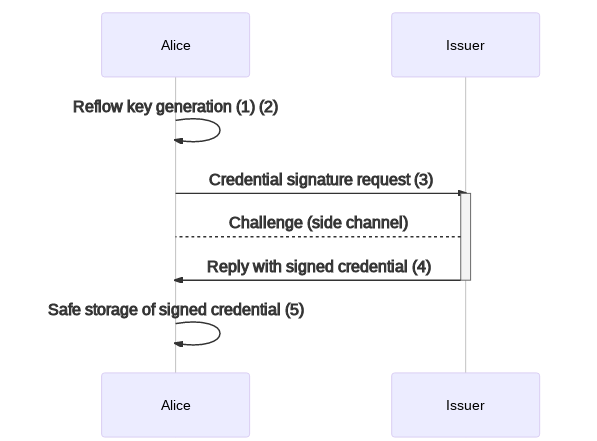}
\end{figure}

Figure~\ref{fig:keygen} illustrates how this process unfolds in a
sequence of communication. An interactive exchange takes place between
the Signer and the Issuer that verifies the possession of the secret
Reflow key and signs a credential to witness this condition.

The public key of the Issuer which is used to sign the credential
should have been public and known by the Signer at the beginning of
the Keygen process.

This code is executed in multiple steps by the Zencode utterances:

\begin{enumerate}

\item \textbf{When I create the reflow key}
  \\ \textbf{and I create the credential key}

  will create inside the \emph{keys} data structure two new secret
  keys: the \emph{credential} and the \emph{bls} keys both consisting
  of \emph{BIG} integers. Secret keys are randomically chosen
  uniformly in $\mathbb{F}_n$, where $n$ is the order of the group
  $\mathbb{G}_1$.

\item \textbf{When I create the credential request}

  will use the \emph{secret keys} to create a new \emph{credential
    request} consisting of a $\lambda$ containing a zero-knowledge
  proof $\pi_s$ of the authenticity of Signer $i$ owning keys
  $(sk_i, PK_i)$ and requesting the Issuer to sign its secret
  attribute $ck_i\in\mathbb{Z}$. The Issuer may request additional
  communications held on a \emph{side channel} in order to establish
  proof-of-possession and the authenticity of the Signer, for instance
  a challenge session in which a random message is demonstrably signed
  by the secret credential key using an El-Gamal signature
  \citep{elgamal} and verified by the public key.

\item \textbf{Given I have a 'credential request'}
 \\ \textbf{When I create the credential signature}
 \\ \textbf{and I create the issuer public key}

  will be executed by the Issuer to sign the credential after the
  proof-of-possession challenge is positive. It consists of a
  $\tilde{\sigma}$ signed used the Issuer's own secret key $sk$ after
  verifying the proof $\pi_s$ inside the received $\lambda$.

\item \textbf{Given I have a 'credential signature'}
 \\ \textbf{When I create the credentials}

 will be executed by the Signer to aggregate one or more
 \emph{credential signature} into the secret \emph{keys} structure.
 The \emph{credentials} consist of a $\sigma_{ck}$ that should be
 stored locally and secretly as it will be later required to create
 the zero-knowledge proof components $\theta$ and $\zeta$ that
 authenticate the Reflow signature against rogue-key attacks.
\end{enumerate}

\onecolumn

\subsection{Signature}


The Reflow signature consists into a multi-signature that can be
indicted by anyone who selects multiple participants. It requires no
interaction between participants and it results into what we call a
\emph{Seal} collecting all the signatures.


Both the credential to sign and the signatures in the reflow seal are
non traceable. The credentials are zero-knowledge proof of possessions
bound to the seal identifier; the public keys of participants are
blinded and known only by the one who indicts the signature, who may
or may not be among participants.

The signatures are non-interactive and made in a multi-party
computation setup: any participant with a credential to sign can do so
without interacting with others, just by adding the reflow signature
to the reflow seal.

Every signature in a seal produces a \emph{fingerprint} unique to the
seal and the signer consisting in a $\zeta$ that is also non
traceable. Using fingerprints a check may be enforced to avoid double
signing and invalidate the reflow seal.

\paragraph{When a seal is created} a session is opened for multiple
participants to add their signature, anyone opening such a process
needs to have an \emph{array of public keys} of all desired
participants.

\begin{lstlisting}[style=zencode,caption={Create reflow seal}]
Scenario reflow: create a new signing session (seal)
Given I have a 'reflow public key array' named 'public keys'
and I have a 'string dictionary' named 'Event'
When I aggregate the reflow public key from array 'public keys'
and I create the reflow identity of 'Event'
and I create the reflow seal with identity 'reflow identity'
Then print the 'reflow seal'
\end{lstlisting}

\paragraph{When a seal is signed} there are two different steps to be
performed: one by the participant publishing the signature and one by
anyone aggregating the signature to the seal. This is necessary since
we are in a multi-party computation session where each step forward in
the process can be performed independently and without requiring
anyone to disclose any secret.

\begin{lstlisting}[style=zencode,caption={Create a reflow signature}]
Scenario reflow: participant signs a seal by producing a signature
Given I am 'Alice'
and I have the 'credentials'
and I have the 'keys'
and I have a 'reflow seal'
and I have a 'issuer public key' from 'The Issuer'
When I create the reflow signature
Then print the 'reflow signature'
\end{lstlisting}

\begin{lstlisting}[style=zencode,caption={Add a signature to a seal}]
Scenario reflow
Given I have a 'reflow seal'
and I have a 'issuer public key' in 'The Authority'
and I have a 'reflow signature'
When I aggregate all the issuer public keys
and I verify the reflow signature credential
and I check the reflow signature fingerprint is new
and I add the reflow fingerprint to the reflow seal
and I add the reflow signature to the reflow seal
Then print the 'reflow seal'
\end{lstlisting}

The session may be considered terminated at any moment by anyone: the
seal can be validated in any moment as-is and it will simply return
false validation if signatures are wrong or incomplete.

\begin{lstlisting}[style=zencode,caption={Compare reflow identities}]
Scenario reflow: compare two different objects using reflow IDs
Given I have a 'reflow seal'
Given I have a 'string dictionary' named Event
When I create the reflow identity of 'Event'
When I rename the 'reflow identity' to 'EventIdentity'
When I verify 'EventIdentity' is equal to 'identity' in 'reflow seal'
Then print the string 'The seal really belongs to this Event'
\end{lstlisting}

\begin{lstlisting}[style=zencode,caption={Verify a seal}]
Scenario reflow: verify a seal contains all valid signatures
Given I have a 'reflow seal'
When I verify the reflow seal is valid
Then print the string 'SUCCESS'
and print the 'reflow seal'
\end{lstlisting}

At this point the seal may be marked as closed as the session is
completed succesfully and all correct signatures have been collected
without duplicates. In case this optional status will be attributed to
the seal then is also possible to remove the \emph{fingerprints}
array from it by reducing its size.

\subsection{Material Passport}

The Reflow material passport consists in a REA model track-and-trace
\citep{trackandtrace} listing. It aggregates Reflow seals by means of
collecting their signed \verb!UID! (unique identifier). The list of
fingerprints in a material passport is not to be confused with that of
a seal: it is the list of all \verb!UID! being traced in one seal,
which adds its own \verb!UID! resulting from the hash to point of the
new message.

The complexity of this implementation requires the adoption of a small
ontology, the Valueflows vocabulary \citep{valueflows}, that briefly
distinguishes between roles of the nodes represented and organizes
their interaction.

A seal may still be signed by one or more participants which are the
\emph{Agents} in the ValueFlows REA model. Agents may be signing a new
\emph{Event} or \emph{Process} alone or toghether with more agents.
All signatures will still be blinded and non-traceable to satisfy
Reflow's privacy requirements, while the \verb!UID! composing the
material passport may be listed in a fingerprint section to
authenticate the track-and-trace graph and perform deeper verification
of all nodes.

\begin{lstlisting}[style=zencode,caption={Create a new material passport}]
Scenario reflow
Given I am 'Alice'
and I have the 'keys'
and I have the 'credentials'
and I have a 'issuer public key' in 'The Authority'
and I have a 'reflow identity'
and I have a 'string dictionary' named 'EconomicEvent'
When I create the material passport of 'EconomicEvent'
Then print the 'EconomicEvent'
and print the 'material passport'
\end{lstlisting}

What follows is the representation of the output, which shows how a
material passport is formatted, keeping the same variable names used
in the mathematical formulas and indicating the type of members ("ecp"
stands for elliptic curve point and a 2 is added to indicate those on a twisted curve) and their size in bytes.

\begin{lstlisting}[caption={Example material passport data}]
material_passport = {
    proof = {
        kappa = ecp2[192] ,
        nu = ecp[49] ,
        pi_v = {
            c = int[32] ,
            rm = int[32] ,
            rr = int[32] 
        },
        sigma_prime = {
            h_prime = ecp[49] ,
            s_prime = ecp[49] 
        }
    },
    seal = {
        SM = ecp[49] ,
        fingerprints = [1] <table 1>,
        identity = ecp[49] ,
        verifier = ecp2[192] 
    },
    zeta = ecp[49] 
}
\end{lstlisting}

As hinted by the optional presence of a list of fingerprints, as well
optional record of the agent's credential \emph{Proof} inside the
material passport, this implementation is extremely flexible and opens
up a range of possible solutions in the field of privacy-enhanced
graph authentication, allowing import and export operations between
federated graph data bases, also considering each data silo may be an
\emph{Issuer} publicly identified and granting agent credentials.

The verification of a material passport implicitly verifies that the
the hash to point of the objects's contents result in the same
\emph{reflow identity}, then executes a cryptographic verification
that the given material passport is the valid signature of that
identity:

\begin{lstlisting}[style=zencode,caption={Verify a material passport}]
Scenario reflow
Given I have a 'string dictionary' named 'EconomicEvent'
and I have a 'issuer public key' in 'The Authority'
and I have a 'material passport'
When I verify the material passport of 'EconomicEvent'
Then print the string 'Valid Event material passport'
\end{lstlisting}

To avoid confusion is important to keep in mind that identities are
\verb!UID!s and fingerprints are \verb!zeta!s.

The simple signature and verification of a single object is not enough
to implement a material passport, we need to authenticate the
\emph{track and trace} of it eventually validating all the complex
history of transformations that lead to the current object. In order
to do so our implementation exploits the aggregation property of
elliptic curve points by making a sum of all object identities present
in the track and trace. Following the ValueFlows logic then, every
Event or Process identity will result from the sum of all identities
of previous Agents, Events and Processes. By recalculating this sum
and checking its signature then one can authenticate the integrity of
the track and trace.

For example, assuming we have an array of \verb!seal! objects found in
material passports of a track and trace result of one level depth:

\begin{lstlisting}[caption={Example material passport data}]
Seals = [2] { {
  SM = ecp[49] ,
  fingerprints = [1] { ecp[49]  },
  identity = ecp[49] ,
  verifier = ecp2[192]
}, {
  SM = ecp[49] ,
  fingerprints = [1] { ecp[49]  },
  identity = ecp[49] ,
  verifier = ecp2[192]
} }
\end{lstlisting}

Then we can calculate a new identity using aggregation (sum value) with
Zencode statements:

\begin{lstlisting}[style=zencode,caption={Aggregate reflow seal identity}]
Given I have a 'reflow seal array' named 'Seals'
When I create the sum value 'identity' for dictionaries in 'Seals'
Then print the 'sum value'
\end{lstlisting}

Which will output an elliptic curve point corresponding to the sum of
all identity elements found in the "Seals" array. The fact that the
verification happens to a sum that cascades down aggregating further
depth level of parent objects allows to modularize the track-and-trace
implementation making its depth configurable.

\twocolumn

\section{Evaluation}
\label{sec:evaluation}

The goal of this section is to produce benchmarks of the implementation of the cryptographic flow, in  realistic conditions of use, so in a way that is similar to how a software solution would use the software. 

Our approach was opposite to testing single algorithms in a sandbox or a profiler, instead we tested Zenroom scripts, written in Zencode, that include also loading and parsing input data from streams or file system, as well as producing output as a deterministically formed JSON object. 

\subsection*{Platforms}
The three target platforms used for the benchmarks were: 
\begin{itemize}
  \item   \textbf{X86-64} on Intel(R) Core(TM) i5-5300U CPU @ 2.30GHz, running Ubuntu 18.04 (64 bit)
  \item   \textbf{ARM 32bit} on \textbf{Raspberry Pi 4} board, running Raspberry Pi OS (32bit, kernel version: 5.10)
  \item  \textbf{ARM 32bit} on \textbf{Raspberry Pi 0} board, running Raspberry Pi OS (32bit, kernel version: 5.10)
\end{itemize}

The the X86-64 machine runs on a Lenovo X250 laptop, with 8GB of RAM.  
The Raspberry 0 and 4 have been chosen as benchmarking platforms because of their similarities to, respectively, very low-cost IoT devices (5 USD) and very low cost mobile phones (sub 100 USD).

\subsection*{Builds} 

Being a self-contained application written in C, Zenroom can be built for several CPU architectures and operative systems, both as command line interface (CLI) application and as a library. The configurations we chose for this benchmark are: 

\begin{itemize}
\item Two binary Zenroom CLI, compiled using the GCC tool-chain in native 32bit ELF binaries, once for the X86-64 platform and once for the ARMv7-32bit. We refer to these builds as \textbf{Zenroom CLI}. 

\item One mixed library, trans-compiled to WASM using the Emscripten tool-chain \citep{zakai2011emscripten}, then built into an NPM package along with a JavaScript wrapper. After trans-compilation, the WASM library is converted to base64 and embedded inside the JavaScript wrapper, which unpacks it at run-time. 
The library runs in browsers (using the native WASM support, currently present in Chromium/Chrome, Firefox and Safari), as well in node JS based application. We refer to this build as \textbf{Zenroom WASM}. 
\end{itemize}

\subsection*{CLI and WASM use cases} 

The use case for Zenroom CLI is typically a server-side application or micro-service. The WASM library can run in the browser, in the front end of a web application or in a server-side application (running on node JS) or a mobile application.

Zenroom can also be built as a native library for Android and iOS (on ARM 32bit, ARM 32bit or X86), but no benchmarks of the Reflow cryptographic flow have been performed using Zenroom as a native library. 
In previous benchmarks, we noted similar performances for the native CLI and the native library versions of Zenroom, we therefore assume that performances would be comparable between the CLI builds we are using and the native library builds.

\subsection*{Testing}
Benchmarking of Zenroom CLI and Zenroom WASM required using two different tools. Both tools performed chained execution of Zencode scripts that would simulate the whole cryptographic flow. Furthermore, the tool allowed to configure the amount of participants in the flow and the repeat the run of selected scripts in order to gather more data.  

\begin{itemize}
\item The testing of Zenroom CLI was executed using a single, self-contained \href{https://github.com/dyne/Zenroom/blob/master/test/zencode_reflow/run-recursive.sh}{bash script}. The script outputs the duration of the execution of each Zenroom scripts, along with the memory usage and size of output data, in a CSV formatted summary. The script also saves to files both the data and scripts for quality control reference.

\item The testing of Zenroom WASM required building a \href{https://github.com/dyne/Reflow-Zencode-WASM-benchmark}{JavaScript application}, running in node JS. The application outputs the duration of the execution of each Zenroom scripts, along with the memory usage, it calculate averages and returns output in a JSON formatted summary.
\end{itemize}

Benchmarks for the cryptographic flow were run on both the CLI and the WASM builds of Zenroom, for each platform, for a total of six different data collections. 

\subsection*{Scripts description}

Our benchmarks provide measurements for the all the steps in the signature flow: for each of them we reported the time of execution (expressed in seconds) and the size of output (expressed in bytes) in different conditions with a progressive number of participants (5, 10, 50, 100, 1000 signatures).

The Zencode scripts used in the flow and for bench-marking can be divided in three categories: 

\begin{itemize}
\item Scripts with \textbf{variable execution times} executed by anyone, marked with (A): they include the creation of the Reflow seal (Session start), the aggregation of the signatures (Collect sign) and the verification of the signatures on Reflow seal. Their duration is proportional to the amount of participants and can differ greatly. These scripts are the most calculation intensive and expected to typically run server-side. The (A) stands for "Anyone" as they don't need an identity or a keypair to run. These scripts are typically expected to run server side.

\item Scripts executed by each participant, marked with (P): they include participant's setup and signing, their duration is not correlated with the amount of participants. These scripts are expected to typically run in the browser or on mobile devices.

\item Scripts executed by the issuer, marked with (I): they include issuer's setup and signing, their duration is not correlated with the amount of participants. These scripts are expected to typically run server-side.
\end{itemize}

Each script was executed 50 times, on each platform and each configuration, and the average execution time was extracted.

\subsection*{Findings}

\begin{itemize}
\item The execution time of the scripts in the (A) group, grows linearly with the amount of participants. 

\item On a X86-64 machine, the benchmarks execute on Zenroom CLI have a comparable duration with their counterparts run on WASM. On the other hand, the execution times differ greatly on the ARM 32bit based Raspberry Pi 0 and 4 machine, when comparing Zenroom CLI and Zenroom WASM executions. 
While investigating the cause of the differences is beyond the purpose of this paper, we speculate this difference signals different levels of optimization, between the of WASM interpreters built for X86-64 on one hand, and ARM 32bit on the other.  

\item The most numerically outstanding benchmark is the execution time of the issuer's keygen script on Raspberry Pi 0, in comparison with the other platforms. Once more, investigating the reason for this discrepancy is beyond the scope of this paper: we do anyway speculate that the ECP2 pairing performed in the script, may justify the large difference.
\end{itemize}

\subsection*{Benchmark conclusions} 

The ultimate goal of this benchmarks is to assess what parts of cryptographic flow are suitable for execution on each platform, both in terms of execution times as well as RAM usage. Concerning the \textbf{RAM usage}, the highest value recorded is in the 5 megabytes range, which we consider to be well within the acceptable range for any of the platforms used in the benchmarks. Future analysis will investigate the possibility to execute the less resource intense parts of the flow, on ultra low power chips such as the \textbf{Cortex M4}, for which a port of Zenroom is available.

Concerning the \textbf{execution time}, if we choose that the longest execution time acceptable for any script of the flow, is up to 1 second, we find that: 

\begin{itemize}
\item Every script of the groups (P) and (I) have acceptable performances on any platform when using Zenroom compiled as CLI. 

\item With a few exceptions, we find a similar result when running scripts of the  (P) and (I) groups, for Zenroom compiled in WASM. With this setup, only three scripts run over the 1 second limit, only on Raspberry Pi 0.
 
\item The scripts in the group (A), when using Zenroom CLI, can be executed within one second for an amount of participants between 100 and 500 on the X86 machine and for a similar amount on a Raspberry Pi 4, while on a Raspberry Pi 0 the limit is between 10 and 50 participants. 

\item The scripts in the group (A), when using Zenroom WASM, can be executed within one second, for an amount of participants between 100 and 500 on the X86 machine and up to 100 on a Raspberry Pi 4, while on a Raspberry Pi 0 the limit is between 2 and 5 participants. 

\end{itemize}



\onecolumn

\section{Benchmarks}
\label{sec:benchmarks}

Following are the benchmarks of the benchmarks for the (A) group of scripts (whose execution times change based on the amount of participants), running in CLI binaries, grouped by platform.

\subsection*{Zenroom CLI X86-64}

\begin{table}[h!]
  \begin{center}
    \caption{Execution timings on \textbf{CLI X64} in $seconds$ per $participant$}
      \label{tab:table1}
        \begin{tabular} {c|c|c|c|c|c|c}
          \toprule
           \textbf{$_S / _P$} & \textbf{2} & \textbf{10} & \textbf{50} & \textbf{100} & \textbf{1000} & \textbf{5000} \\
          \midrule
          (A) Session start & 0.0188 & 0.0253 & 0.0410 & 0.0623 & 0.1056 & 0.5039 \\
          (A) Collect sign & 0.0729 & 0.0730 & 0.1031 & 0.1373 & 0.7435 & 3.6061 \\
          (A) Verify sign & 0.0405 & 0.0515 & 0.0996 & 0.1619 & 1.3974 & 7.0460 \\
      \bottomrule 
    \end{tabular}
  \end{center}
\end{table}

\begin{figure}[h!]
    \centering
    \includegraphics[width=4in, height=2.6in]{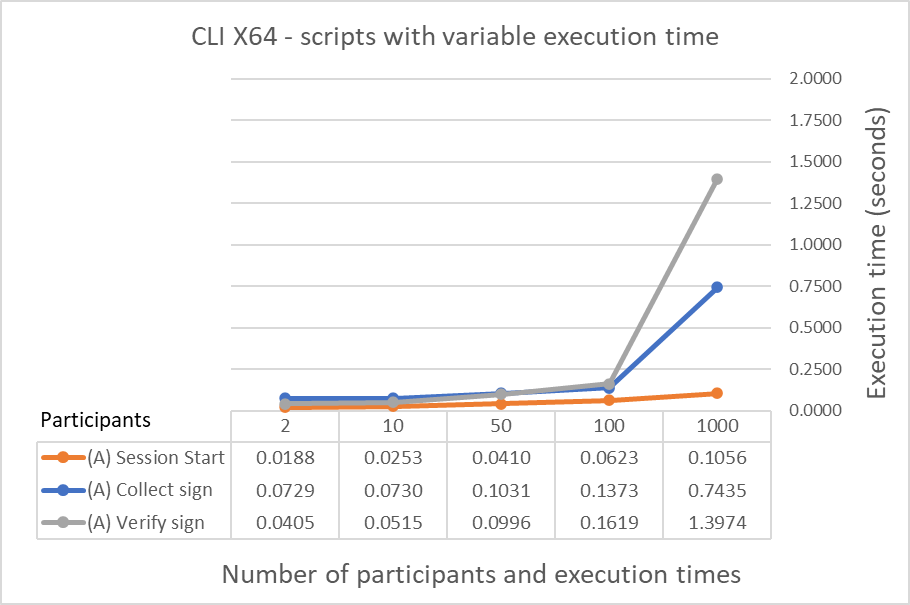}
    \label{fig:galaxy}
\end{figure}

\subsection*{Zenroom CLI Raspberry Pi 4}

\begin{table}[h!]
  \begin{center}
    \caption{Execution times on \textbf{CLI Raspberry 4} in $seconds$ per $participant$}
      \label{tab:table1}
        \begin{tabular} {c|c|c|c|c|c}
          \toprule
           \textbf{$_S / _P$} & \textbf{2} & \textbf{10} & \textbf{50} & \textbf{100} & \textbf{1000} \\
          \midrule
          (A) Session start & 0.0516 & 0.0602 & 0.0674 & 0.0750 & 0.2762 \\
          (A) Collect sign & 0.2087 & 0.2273 & 0.3031 & 0.3942 & 2.1849 \\
          (A) Verify sign & 0.1049 & 0.1393 & 0.2978 & 0.4831 & 4.0688 \\
      \bottomrule 
    \end{tabular}
  \end{center}
\end{table}

\begin{figure}[h!]
    \centering
    \includegraphics[width=4in, height=2.6in]{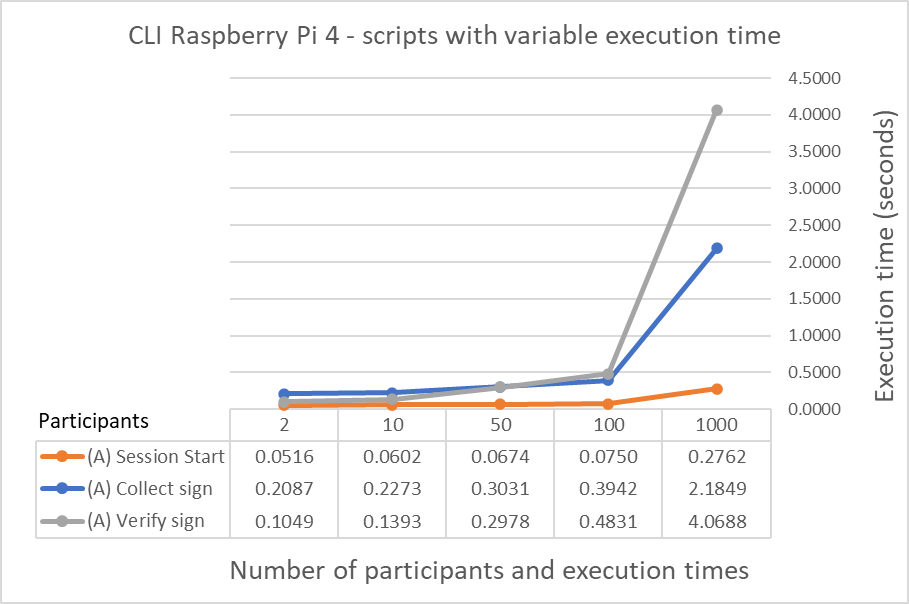}
    \label{fig:galaxy}
\end{figure}

\pagebreak
\newpage

\subsection*{Zenroom CLI Raspberry Pi 0}

\begin{table}[h!]
  \begin{center}
    \caption{Execution times for\textbf{Zenroom CLI Raspberry Pi 0} in $seconds$ per $participant$}
      \label{tab:table1}
        \begin{tabular} {c|c|c|c|c|c}
          \toprule
           \textbf{$_S / _P$} & \textbf{2} & \textbf{10} & \textbf{50} & \textbf{100} & \textbf{1000} \\
          \midrule
          (A) Session start & 0.2760 & 0.2936 & 0.3413 & 0.3725 & 1.3097 \\
          (A) Collect sign & 1.0439 & 1.1458 & 1.3688 & 1.7669 & 9.6403 \\
          (A) Verify sign & 0.5753 & 0.7199 & 1.3381 & 2.1971 & 17.8956 \\
      \bottomrule 
    \end{tabular}
  \end{center}
\end{table}

\begin{figure}[h!]
    \centering
    \includegraphics[width=4in, height=2.6in]{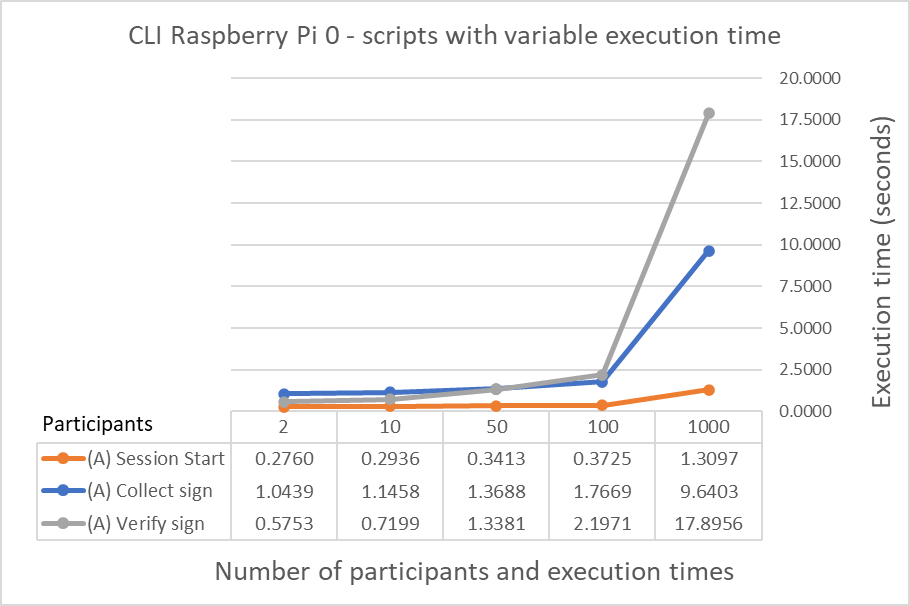}
    \label{fig:galaxy}
\end{figure}

\subsection*{Zenroom CLI - all platforms}

Following are the benchmarks of the benchmarks for the (P) and the (I) groups of scripts (whose execution times don't change based on the amount of participants), with a comparison of all the platforms, running in CLI binaries.


\begin{table}[h!]
  \begin{center}
    \caption{Execution timings on \textbf{CLI} of scripts in $seconds$ per $platform$}
      \label{tab:table1}
        \begin{tabular} {c|c|c|c|c|c}
          \toprule
\textbf{$script / platf.$} & \textbf{(P) Keygen} & \textbf{(P) Request} & \textbf{(P) Pub-Key} & \textbf{(P) Aggr. Cred.} & \textbf{(P) Sign Session} \\
          \midrule
			X86-X64	&	0.0137	&	0.0391	&	0.0196	&	0.0219	&	0.0521		\\
Raspberry Pi 4	&	0.0373	&	0.0883	&	0.0475	&	0.0579	&	0.1456		\\
Raspberry Pi  0	&	0.2492	&	0.5165	&	0.3017	&	0.3469	&	0.6603		\\
      \bottomrule 
    \end{tabular}
  \end{center}
\end{table}

\begin{table}[h!]
  \begin{center}
    \caption{Execution timings on \textbf{CLI} of scripts in $seconds$ per $platform$}
      \label{tab:table1}
        \begin{tabular} {c|c|c|c}
          \toprule
\textbf{$script / arch$} & \textbf{(I) Keygen} & \textbf{(I) Pub-Key} & \textbf{(I) Sign Req.} \\
          \midrule
			X64	&	0.0129	&	0.0223	&	0.0524	\\
Raspberry Pi 4	&	0.0404	&	0.0597	&	0.1405	\\
Raspberry Pi  0	&	0.2467	&	0.3766	&	0.7687	\\
      \bottomrule 
    \end{tabular}
  \end{center}
\end{table}

\begin{figure}[h!]
    \centering
    \includegraphics[width=6in, height=2.6in]{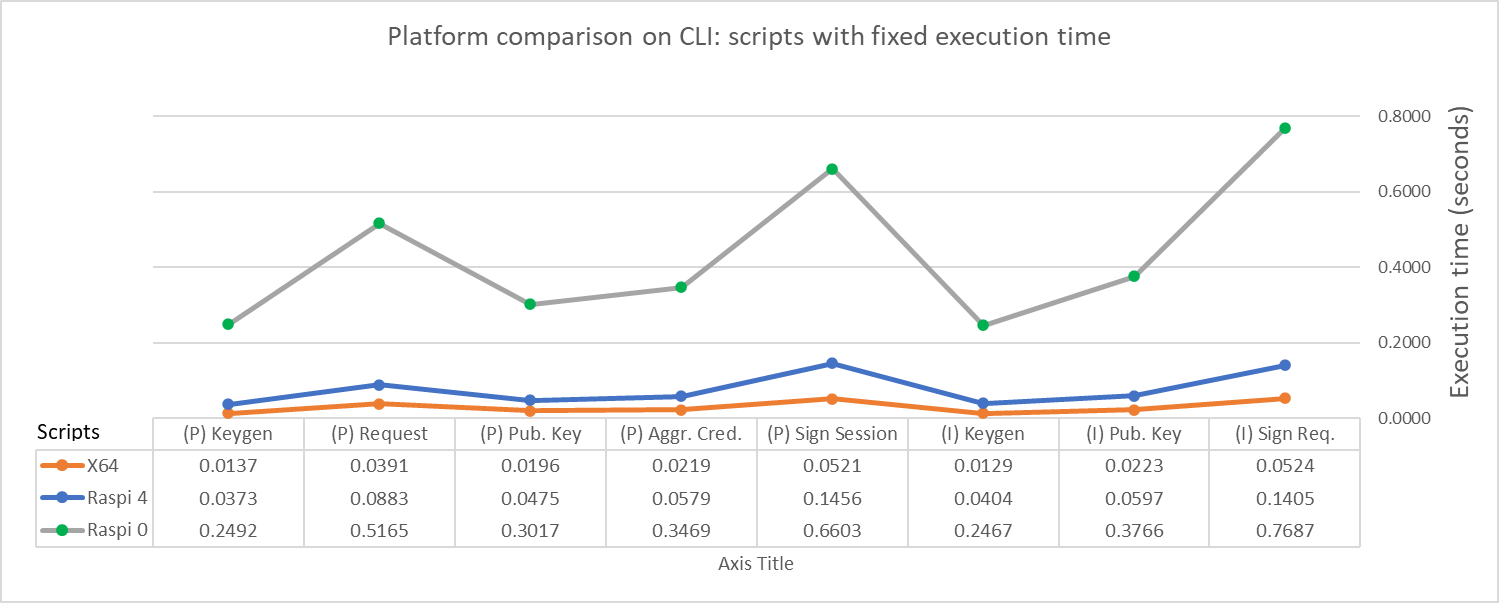}
    \label{fig:galaxy}
\end{figure}

\newpage
\pagebreak

Following are the benchmarks of the benchmarks for the (A) group of scripts (whose execution times change based on the amount of participants), running in WASM libraries, grouped by platform.

\subsection*{Zenroom WASM X86-64}

\begin{table}[h!]
  \begin{center}
    \caption{Execution timings on \textbf{WASM X64} in $seconds$ per $participant$}
      \label{tab:table1}
        \begin{tabular} {c|c|c|c|c|c}
          \toprule
           \textbf{$_S / _P$} & \textbf{2} & \textbf{10} & \textbf{50} & \textbf{100} & \textbf{1000} \\
          \midrule
          (A) Session start & 0.0324 & 0.0312 & 0.0373 & 0.0516 & 0.1749 \\
          (A) Collect sign & 0.1180 & 0.1092 & 0.1510 & 0.2467 & 1.1563 \\
          (A) Verify sign & 0.0695 & 0.0688 & 0.1495 & 0.2641 & 2.1554 \\
      \bottomrule 
    \end{tabular}
  \end{center}
\end{table}

\begin{figure}[h!]
    \centering
    \includegraphics[width=4in, height=2.6in]{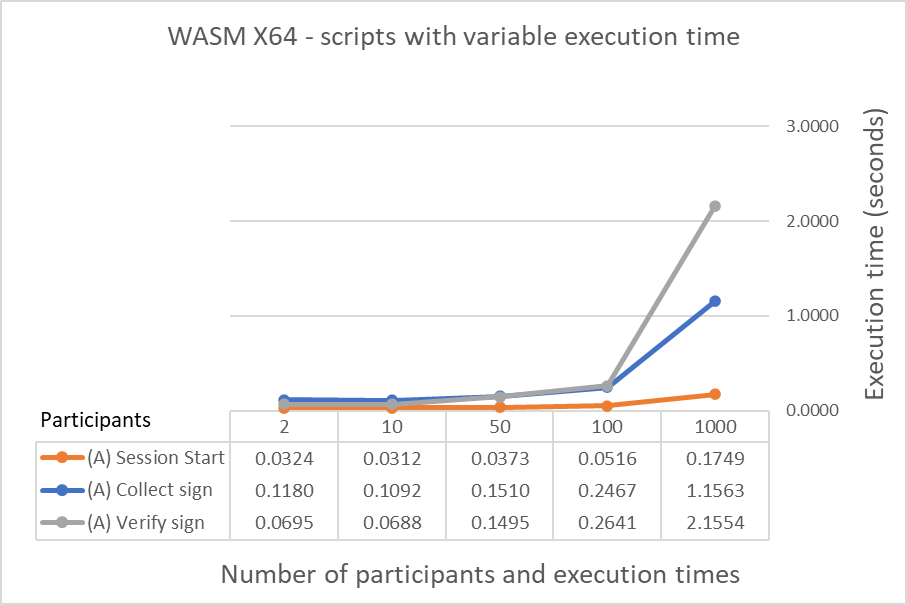}
    \label{fig:galaxy}
\end{figure}

\newpage

\subsection*{Zenroom WASM Rasperry Pi 4}

\begin{table}[h!]
  \begin{center}
    \caption{Execution timings on \textbf{WASM Raspberry Pi 4} in $seconds$ per $participant$}
      \label{tab:table1}
        \begin{tabular} {c|c|c|c|c|c}
          \toprule
           \textbf{$_S / _P$} & \textbf{2} & \textbf{10} & \textbf{50} & \textbf{100} & \textbf{1000} \\
          \midrule
          (A) Session start & 0.1065 & 0.1092 & 0.1269 & 0.1459 & 0.5432 \\
          (A) Collect sign & 0.3897 & 0.4170 & 0.5741 & 0.7654 & 4.3594 \\
          (A) Verify sign & 0.1880 & 0.2510 & 0.5621 & 0.9494 & 8.1078 \\
      \bottomrule 
    \end{tabular}
  \end{center}
\end{table}

\begin{figure}[h!]
    \centering
    \includegraphics[width=4in, height=2.6in]{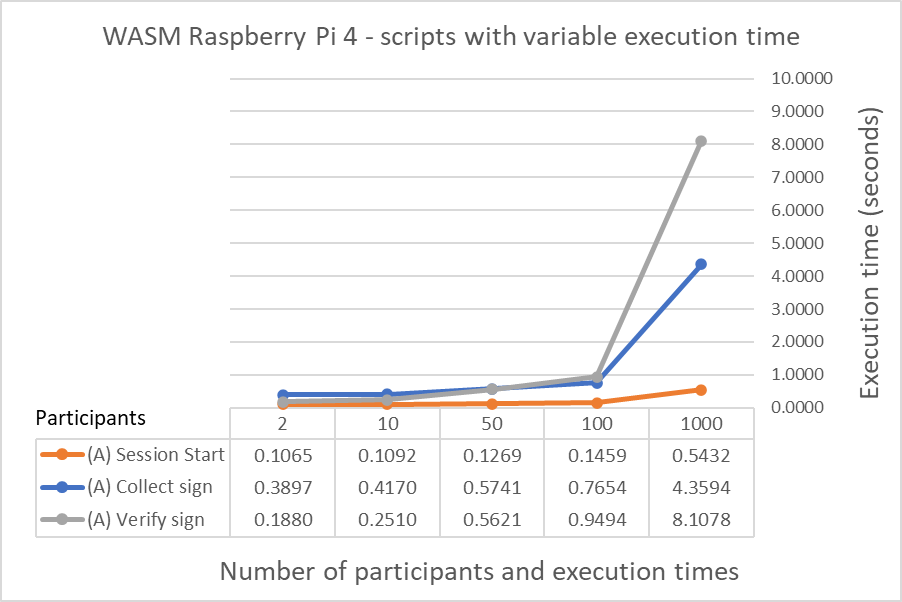}
    \label{fig:galaxy}
\end{figure}

\subsection*{Zenroom WASM Rasperry Pi 0}

\begin{table}[h!]
  \begin{center}
    \caption{Execution timings on \textbf{WASM Raspberry Pi 0} in $seconds$ per $participant$}
      \label{tab:table1}
        \begin{tabular} {c|c|c|c|c|c}
          \toprule
           \textbf{$_S / _P$} & \textbf{2} & \textbf{10} & \textbf{50} & \textbf{100} & \textbf{1000} \\
          \midrule
          (A) Session start & 0.8688 & 0.8711 & 1.0392 & 1.2257 & 4.7067 \\
          (A) Collect sign & 2.7272 & 2.8665 & 3.9580 & 5.1851 & 29.0912 \\
          (A) Verify sign & 1.3818 & 1.7820 & 3.8503 & 6.3604 & 53.8938 \\
      \bottomrule 
    \end{tabular}
  \end{center}
\end{table}

\begin{figure}[h!]
    \centering
    \includegraphics[width=4in, height=2.6in]{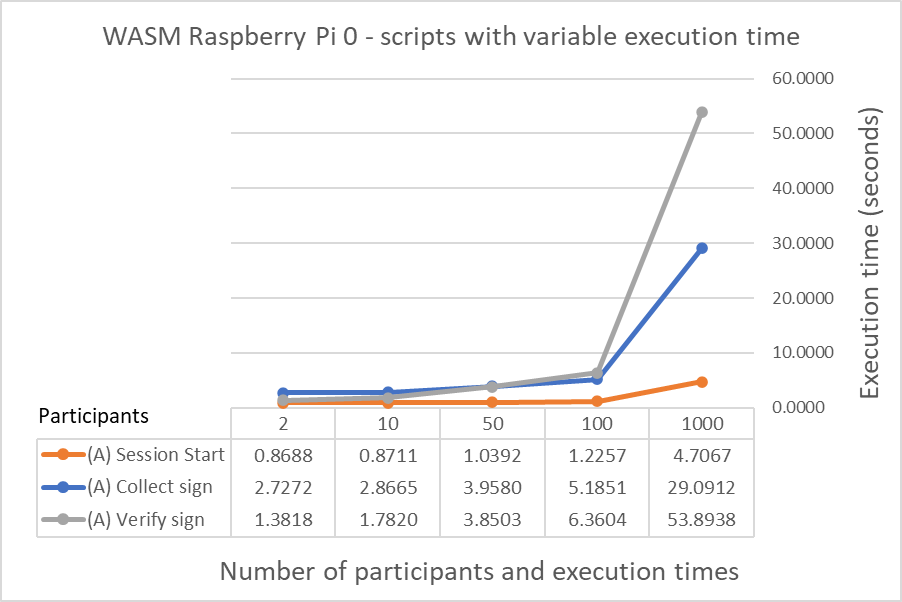}
    \label{fig:galaxy}
\end{figure}

\newpage

\subsection*{Zenroom WASM - all platforms}

Following are the benchmarks of the benchmarks for the (P) and the (I) groups of scripts (whose executions don't change based on the amount of participants), with a comparison of all the platforms, running in CLI binaries.


\begin{table}[h!]
  \begin{center}
    \caption{Execution times on \textbf{Zenroom WASM} of scripts in $seconds$ per $platform$}
      \label{tab:table1}
        \begin{tabular} {c|c|c|c|c|c}
          \toprule
\textbf{$script / platf.$} & \textbf{(P) Keygen} & \textbf{(P) Request} & \textbf{(P) Pub-Key} & \textbf{(P) Aggr. Cred.} & \textbf{(P) Sign Session} \\
          \midrule
X86-X64			&	0.0213	&	0.0490	&	0.0262	&	0.0304	&	0.0745		\\
Raspberry Pi  4	&	0.169	&	0.3365	&	0.2095	&	0.173	&	0.2705		\\
Raspberry Pi  0	&	0.6275	&	1.291	&	0.7185	&	0.811	&	1.9435		\\
      \bottomrule 
    \end{tabular}
  \end{center}
\end{table}

\begin{table}[h!]
  \begin{center}
    \caption{Execution times on \textbf{Zenroom WASM} of scripts in $seconds$ per $platform$}
      \label{tab:table1}
        \begin{tabular} {c|c|c|c}
          \toprule
\textbf{$script / platf$} &  \textbf{(I) Keygen}& \textbf{(I) Pub-Key}& \textbf{(I) Sign Req.} \\
          \midrule
X86-X64				&		0.0204	&	0.0317	&	0.0746	\\
Raspberry Pi  4	&		1.01	&	0.276	&	0.6585	\\
Raspberry Pi  0	&		9.958	&	1.101	&	2.6355	\\
      \bottomrule 
    \end{tabular}
  \end{center}
\end{table}

\begin{figure}[h!]
    \centering
    \includegraphics[width=6in, height=2.6in]{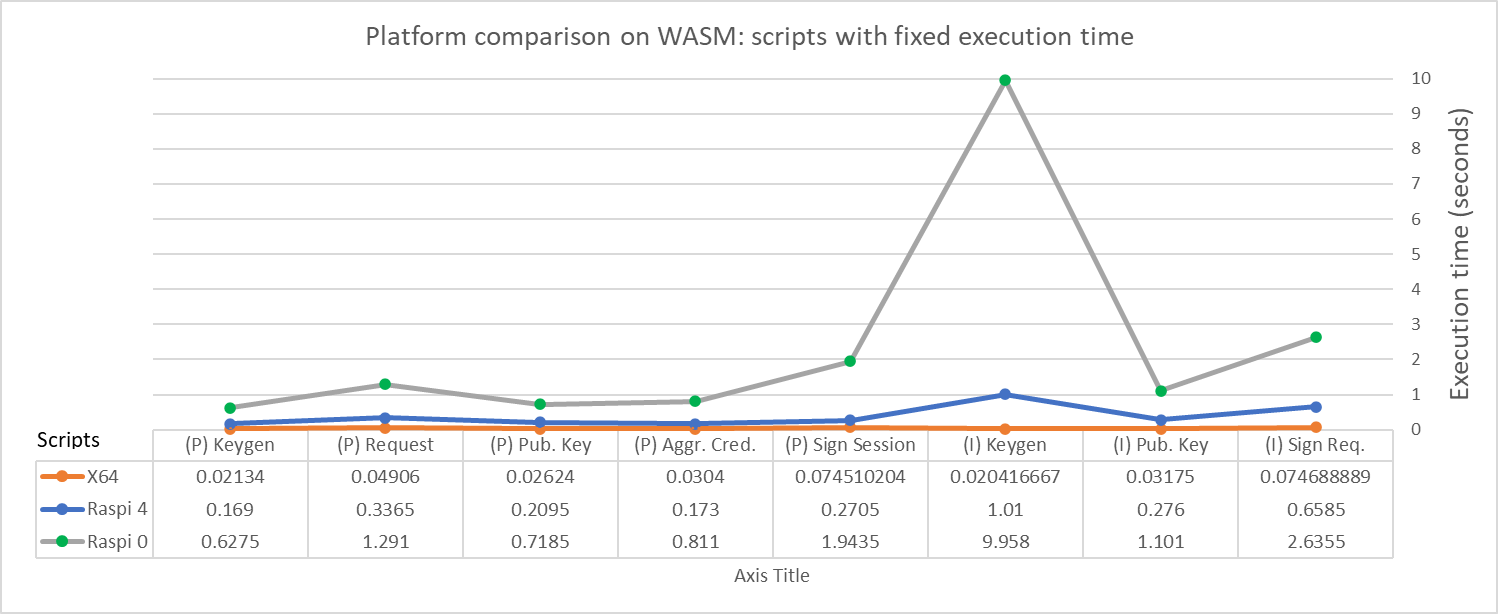}
    \label{fig:galaxy}
\end{figure}

\newpage

\subsection*{RAM usage}

The benchmarking of the RAM has been performed on Zenroom CLI only. The RAM usage showed negligible differences among the X86 and the arm builds. The tables show the results of the X86 benchmarks.   

\begin{table}[h!]
  \begin{center}
    \caption{Execution timings on \textbf{RAM Usage} in $KiloBytes$ per $participant$}
      \label{tab:table1}
        \begin{tabular} {c|c|c|c|c|c|c}
          \toprule
           \textbf{$KB / Participant$} & \textbf{2} & \textbf{10} & \textbf{50} & \textbf{100} & \textbf{1000} & \textbf{5000} \\
          \midrule
            (A) Session start & 500 KB & 507 KB & 534 KB & 589 KB & 1289 KB & 4543 KB \\	
            (A) Collect sign & 520 KB & 548 KB & 665 KB & 817 KB & 1651 KB & 4368 KB	\\		
            (A) Verify sign & 507 KB & 532 KB & 653 KB & 801 KB & 1660 KB & 4695 KB	\\
      \bottomrule 
    \end{tabular}
  \end{center}
\end{table}

\begin{table}[h!]
  \begin{center}
    \caption{Execution times on \textbf{RAM Usage} in $KiloBytes$}
      \label{tab:table1}
        \begin{tabular} {c|c|c|c|c|c}
          \toprule
         & \textbf{(P) Keygen} & \textbf{(P) Request} & \textbf{(P) Pub-Key} & \textbf{(P) Aggr. Cred.} & \textbf{(P) Sign Session} \\
          \midrule
 			& 498 KB	&  512 KB	&  484 KB	&  497 KB	&  541 KB	\\

      \bottomrule 
    \end{tabular}
  \end{center}
\end{table}

\begin{table}[h!]
  \begin{center}
    \caption{Execution times on \textbf{RAM Usage} in $KiloBytes$}
      \label{tab:table1}
        \begin{tabular} {c|c|c|c}
          \toprule
          &  \textbf{(I) Keygen}& \textbf{(I) Pub-Key}& \textbf{(I) Sign Req.} \\
          \midrule
				&	488 KB	&  491 KB	&  504 KB	\\
      \bottomrule 
    \end{tabular}
  \end{center}
\end{table}

\twocolumn

\section{Conclusion}

This article and the referenced free and open source implementation
made in Zencode successfully provides an easy, flexible and well
performant way to realize anonymous signatures made by zero-knowledge
proof credential holders and aggregated in multi-party computation
decentralized environments.

The privacy-preserving features of Reflow signatures are protected
from rogue-key attacks by enforcing credential issuance, at the price
of requiring one or multiple authorities. From this approach it comes
the feature of generating signature "fingerprints" that are contextual
and become traceable only with the consensus of participants (for
instance by replicating the signature). These fingerprints are also
useful to avoid double-signing and to strengthen the track-and-trace
features of the most advanced material passport use-case scenario.

The use-cases we observed haven't required the adoption of multiple
authorities, however it is possible to enhance our scheme by porting a
threshold credential issuance mechanism based on Lagrange
interpolation as described in Coconut's paper \citep{coconut-2018}.

Generally speaking the Reflow cryptographic system can be adapted to
authenticate many sorts of graph data sets; this aspect goes well
beyond the state of the art and its application may comes useful to an
upcoming generation of non-linear distributed ledger technologies.

\subsection{Future directions}

This cryptographic scheme is named after the Reflow project and currently
follows its path to serve the use-case of circular economy projects
for which the integrity of accounting, the portability of data and the
privacy of users are of paramount importance. We are proud of the
priority this cryptography research gives to use-cases dealing with
environmental responsibility and will carry on this ethos through its
future developments.

Standardisation is an important point in order to establish federated
and decentralised use of Reflow across a variety of contexts, so we have
high incentives in integrating this signature in W3C standards as well
submitting it to the attention of the Object Management Group.

As hinted in the overview, the discourse around disposable identities
is also very interesting and deserves further interaction as it may
open up more privacy preserving possibilities of development in the
field of health-care and transportation as well public services and
accounting traceability for KYC/AML practices.

The possibility to aggregate Reflow signatures may also be exploited
to serve need-to-know schemes where different actors may be required
to sign different parts of a document without knowing the whole until
its final publication, a scenario present in peer-review processes
that may improve their impartiality.

Being at the core of the Reflow project development of a federated
database of information that serves the piloted circular-economy
use-cases, the main deployement of the Reflow crypto model is inside the
Zenpub software component and will refine its interaction to GUI and
dashboard components through a GraphQL interface and ActivityPub event
protocol, as well leveraging the federating capabilities of this scheme
to a stable and replicable implementation.

The developments of next-generation DLT applications as Holochain
\citep{holochain} and Hashgraph \citep{hashgraph} are also very
interesting, in particular Holochain is known have the ValueFlows REA
approach in course of implementation and may soon provide a platform
where to further distribute the graph data beyond the federated server
provision currently implemented in ZenPub.

\subsection{Acknowledgments}

The development and demonstration of the Reflow signature scheme and the
material passport has been partially funded by the European project
"REFLOW" referenced as H2020-EU.3.5.4 with grant nr.820937; the
multi-party signature implementation has been partially funded by Riddle
\& Code GmbH as private commission.

We are grateful to all our colleagues at Dyne.org for the help and
insights they have shared contributing to this work and in particular to
Puria Nafisi Azizi, Danilo Spinella, Adam Burns and Ivan Minutillo; as
well to Thomas Fuerstner for sharing his visions and passion for crypto
development with us; and to prof. Massimiliano Sala and Giancarlo
Rinaldo for facilitating Alberto's stage program with the dept. of
Mathematics at University of Trento. Our gratitude goes also to our
former colleagues in the DECODE project George Danezis and Alberto
Sonnino for creating Coconut, a wonderful crypto scheme on which Reflow
is grafted. Last not least we are grateful to all the REFLOW project
consortium represented by prof. Cristiana Parisi as principal
investigator.

\bibliographystyle{unsrtnat}

\bibliography{references}

\listoffigures

\end{document}